\documentclass[10pt]{iopart}
\usepackage{iopams}
\usepackage{amssymb}
\usepackage{mathrsfs}
\expandafter\let\csname equation*\endcsname\relax			
\expandafter\let\csname endequation*\endcsname\relax		
\usepackage{amsmath} 
\usepackage{float} 									
\usepackage{graphicx}
\usepackage{color}
\usepackage[tight,TABTOPCAP]{subfigure}
\usepackage{epstopdf}
\newcommand{\bt}			{\beta}
\newcommand{\gm}			{\gamma}
\newcommand{\dlt}		{\delta}
\newcommand{\eps}		{\epsilon}
\newcommand{\veps}		{\varepsilon}

\newcommand{\lm}			{\lambda}

\newcommand{\Gm}			{\Gamma}
\newcommand{\Dlt}		{\Delta}

\newcommand{\mc}[1]{\mathcal{ #1}}						
\newcommand{\trm}[1]{\textrm{ #1}}						
\newcommand{\mbb}[1]{\mathbb{ #1}}						

\newcommand{\unl}[1]{\underline{ #1}}					

\newcommand{\ud}{\mathrm{d}} 								

\makeatletter
\newbox\slashbox \setbox\slashbox=\hbox{$/$}
\newbox\Slashbox \setbox\Slashbox=\hbox{\large$/$}
\def\pFMslash#1{\setbox\@tempboxa=\hbox{$#1$}
  \@tempdima=0.5\wd\slashbox \advance\@tempdima 0.5\wd\@tempboxa
  \copy\slashbox \kern-\@tempdima \box\@tempboxa}
\def\pFMSlash#1{\setbox\@tempboxa=\hbox{$#1$}
  \@tempdima=0.5\wd\Slashbox \advance\@tempdima 0.5\wd\@tempboxa
  \copy\Slashbox \kern-\@tempdima \box\@tempboxa}

\def\miss#1{\ifmmode{/\mkern-11mu #1}\else{${/\mkern-11mu #1}$}\fi}

\newcommand{\tbf}[1]{\textbf{#1}}														

\newcommand{\reff}[1]{~(\ref{#1})}															
\newcommand{\citte}[1]{~\cite{#1}}															

\newcommand{\ie}{\emph{i.e.\,}}															
\newcommand{\eg}{\emph{e.g.\,}}														

\numberwithin{equation}{section}				
\usepackage{mathtools}

\begin{document}
\makeatletter


\title[One-loop order effects from one...]{One-loop order effects from one universal extra dimension on $\lm\phi^{4}$ theory}
\author{M. A. L\'{o}pez-Osorio$^{(a)}$, E. Mart\'{i}nez-Pascual$^{(a)}$, G. I. N\'{a}poles-Ca\~{n}edo$^{(b)}$, J. J. Toscano$^{(b)}$}
\address{$ ^{(a)} $Departamento de Ciencias Naturales y Exactas, \mbox{Centro Universitario de los Valles,~Universidad de Guadalajara,}\\
\mbox{Carretera Guadalajara-Ameca Km 45.5, CP 46000, Ameca, Jalisco, M\'exico.}\\ $^{(b)}$Facultad de Ciencias F\'{\i}sico Matem\'aticas,
Benem\'erita Universidad Aut\'onoma de Puebla, Apartado Postal
1152, Puebla, Puebla, M\'exico.\\}
\ead{eric.martinez@academicos.udg.mx}
\vspace{10pt}
\begin{indented}
\item[]February 2020
\end{indented}
\vspace{10pt}
\begin{abstract}
The self-interacting $\lm\phi^{4}$ scalar field theory is a warhorse in quantum field theory. Here we explore the one-loop order impact from one universal extra dimension, $S^{1}/\mathbb{Z}_{2}$, to the self-energy and four point vertex functions associated to this theory. Such effects come as an infinite number of UV divergences corresponding to an infinite superposition of excited KK particles around the loop. We show that dimensional regularisation is adequate enough to control them in terms of the product of the one dimensional inhomogenous Epstein zeta function times the gamma function. From the analytical properties of these functions, the UV divergences are extracted and the counterterms defined; the latter turn out to be of canonical dimension four at the Lagrangian level. We use both, the MS-scheme and a mass-dependent subtraction scheme to remove divergences. Only the latter manifestly satisfy the decoupling theorem.
\end{abstract}


\section{Introduction}
\label{in}
The idea of extra dimensions in field theories dates back to the 1920s\citte{kal21,kle26}, since then, many related proposals, extensions and critical judgements about this idea have been gravitating in theoretical physics\citte{ein36,ker64,tra70,cho75,wit81}. In particular,  phenomenological string theory, naturally formulated in this framework, led the community to revisit the issue decades later from different perspectives. The use of the braneworld scenario allows to consistently lower the typical scale of quantum gravity to TeVs by choosing the number and size of spatial extra dimensions\citte{ark98,lyk96} in the so-called \emph{large} extra dimensional models, in contrast with \emph{warped}\citte{ran99} or \emph{universal}\citte{app01} extra dimensional models; the latter were particularly inspired by\citte{ant94},  and accessible lectures on all these models are for instance\citte{kri04,che09}. Models with universal extra dimensions (UEDs) have recently been a matter of interest since, for example, they are capable of suggesting answers to some still valid questions within the Standard Model (SM), as it is the possible explanation of having three generations of fermions as a result of the global cancellation of anomalies in the SM extension\citte{dobrescu}, in addition, they provide a convincing explanation of proton stability\citte{Ponton}. UED models compactified on $S^{1}/\mbb{Z}_{2}$ orbifolds preserve KK parity and provide a dark matter candidate in the form of stable KK partners\citte{Feng}. Models containing spatial extra dimensions continue to be a topic with phenomenological importance, and searching of this phenomenon in nature is continuously contemplated in the experiments of the Large Hadron Collider\citte{CMS, ATLAS}.

The basic idea in a UED model is rather simple\citte{hoo07}: the stage is a factorisable spacetime geometry with compact spatial extra dimensions, commonly an orbifold, on which a field theory is defined. The dimensional reduction takes place once the extra dimensional content of the fields is harmonically expanded, under certain boundary conditions, and the extra dimensions themselves are integrated out at the action level; the resulting model, which we referred to it as the \emph{effective} or \emph{dimensionally reduced theory}, comprises an infinite number of fields defined on $\mc{M}^{4}$, namely, the zero modes (one for each original field) and an infinite number of Kaluza-Klein (KK) mode fields. These fields originally enter as coefficients in the expansions allowed by the corresponding extra spatial geometry. In the dimensionally reduced theory, the KK fields are the only remaining information from the extra dimensions, hence we attribute any of their contributions on the effective theory as some kind of extra dimensional input. 

In this paper we are interested in the dimensionally reduced theory obtained from the self-interacting $\lm_{(5)}\Phi^{4}$ model, plus all kind of Lorentz invariant operators $\mc{O}^{5+r}(\Phi,\partial_{M}\Phi)$ of canonical dimension greater than five suppressed by some mass scale as in any effective field theory, defined on the five dimensional factorisable spacetime $\mc{M}^{4}\times S^{1}/\mbb{Z}_{2}$. Here $\mc{M}^{4}$ is the Minkowski spacetime. Reducing the dimension, as in the aforementioned sense, makes the field content $\Phi(x,\bar{x})$, with $(x,\bar{x})=(x^{\mu},x^{5})\in \mc{M}^{4}\times S^{1}/\mbb{Z}_{2}$, to be recast into the zero mode $\phi(x)$ and the infinite number of excited KK fields $\phi^{({k})}(x)$, where ${k}$ denotes non-zero Fourier labels. After integrating out the extra dimension, the dimensionally reduced theory consists of an infinite number of fields gathered into four sectors: $(i)$ the pure zero mode sector, which corresponds to the well-known real scalar self-interacting $\lm\phi^{4}$ Lagrangian, $(ii)$ the interaction terms  between the zero mode and the excited KK fields; $(iii)$ the pure KK excited sector, from which the bare propagator of each KK excited field can be read and some quartic interactions among these fields exist; and $(iv)$ the sector that contains all kind of Lorentz invariant operators $\mc{O}^{4+r}(\phi,\phi^{({k})},\partial_{\mu}\phi,\partial_{\mu}\phi^{({k})})$ of canonical dimension greater than $4$ appropriately suppressed by some mass scale such that every sector is of canonical dimension four. Since KK fields actually correspond to Fourier coefficients in the expansion of the field $\Phi(x,\bar{x})$, they define the profile of $\Phi$ in the extra dimension and we interpret their presence as an effective input from the extra dimension, besides they can be interpreted as heavy fields from an effective field theory perspective~\citte{akh08,akh09}. Bearing this in mind, with the appropriate terms in the dimensionally reduced theory, in this paper we calculate the one-loop order contributions from the extra dimension, or excited KK fields, to the two- and four-point vertex functions associated to the $\lm\phi^{4}$ Lagrangian, we regard these functions as standard vertex functions from which the Green's functions with only external light particles can be reconstructed. One of the main challenges in these calculations relies on the infinite number of interactions between the zero mode and the different KK excited modes. In turn, they imply at the one-loop order an infinite number of graphs, each of these contributing with a UV divergence and we will consider all of them. To be more precise, in this scenario, a typical one-order loop amplitude involves a series of UV divergences as $\sum_{k=1}^{\infty}\int\ud^{4}k $ which should be regularised in order to make sense of these type of contributions. In the same spirit of a previous work co-authored by us\citte{QED5d} on QED with extra dimensions, we argue that both the series and each UV divergence can be handled in this purely scalar field theory by means of the well-known dimensional regularisation, whose ultimate justification is found in the analytical continuation of divergent integrals to the complex plane\citte{thooft72}. It turns out that, once the dimensional regularisation is performed, the infinite number of UV divergences results into the one dimensional inhomogeneous Epstein zeta function\citte{eps03, Ep1992,elibook12,Eli1994} times the gamma function. 

In this communication we show, using dimensional regularisation and keeping the infinite number of contributions from excited KK modes, that the extra dimensional effect on the two- and four-point vertex functions of the self-interacting $\lm\phi^{4}$ to the one-loop order can be encapsulated into the product of the one dimensional inhomogeneous Epstein zeta function times the gamma function. Therefore the whole process of extracting UV divergences is reduced to analyse the singularities of such product of functions. Remarkably, the poles of this product can be explicitly obtained in the low energy limit $m^{2}/R^{-2}\ll 1$, where $R$ is the radius of the extra dimension and $m$ is the mass associated to the light field $\phi$. At the one-loop order, we show that in both, the two- and four-point standard vertex functions, the extra dimensional contribution contains divergences of the typical form $(1/\eps)$ with constant factors, where $\eps$ is the usual parameter in dimensional regularisation ($\eps=4-D$).

In order to cancel out the UV divergences already described at the one-loop order we use two different types of schemes to define the counterterms at the Lagrangian level: the MS-scheme and a mass dependent subtraction scheme with a kinematical or subtraction scale $M^{2}$. Interestingly, we will see that in the presence of one UED, with the geometry given above, there is no need to go beyond operators of canonical dimension higher than four to define these counterterms at the one-loop level. The reason for the examination of two different subtraction schemes is to  gain some insight in relation with the decoupling theorem by Appelquist and Carazzone\citte{App75}. It is known that although the MS-scheme removes the UV divergences, the corresponding amplitudes and beta function, do not obey this theorem; in contrast, when using a mass dependent subtraction this theorem is manifestly satisfied\citte{Petrov}. This behaviour is followed in the context of the dimensionally reduced theories as it was already pointed out in\citte{QED5d} and it will be confirmed in the present communication.

As a word of warning, we must say that in this work we define the physical mass and coupling constant by handling the infinite number of UV divergences present at relevant \emph{light particle} vertex functions to the 1-loop approximation, that is, we do not attempt to give a meaningful complete approach to the renormalization of the whole dimensionally reduced model in this work; nevertheless, our contribution may shed some light on this goal.

In terms of\citte{cle02} our method is closer to a `KK-renormalisation', in contrast to the `KK-regularisation' in which the infinite sums over KK modes are performed before the integration over internal loops\citte{del99}. A key mathematical resource in our work is the Epstein zeta function\citte{eps03}, which is a generalisation of the well-known Riemann zeta function\citte{rie58}; both being very interesting objects in their own right. The systematic use in physics of zeta-regularisation methods dates back to the 1970s with seminal works as\citte{dow76} and\citte{haw77} in the context of effective theories and path integrals on curved spacetimes, respectively. Applications of zeta functions can also be spotted in quantum gravity models as well as cosmology\citte{eli00}, string theory\citte{polbook}, and crystallography\citte{zuc74}. Complete modern treatises on this topic are\citte{elibook12, Eli1994}. Probably the most famous application of the zeta-regularisation is found in the Casimir effect\citte{cas48} where the calculation of the physical vacuum energy of a quantised field in the presence of external boundary configurations is performed.  More recently, the computation of the vacuum energy for scalar fields at different temperature limits has been solved by the use of the Epstein zeta function\citte{ede06,kir91}. We mention in passing that the emergence of the Epstein zeta function in our regularisation proposal may be not as surprising as it looks since the boundaries of the compact extra dimensions may resemble the external boundary configurations in the Casimir effect.

The rest of the paper is organised as follows. In Sec.~\ref{secc:5Dmodel} we describe the dimensionally reduced model for the case of one UED, we point which terms are necessary to calculate the contributions from extra dimensions to the $\lambda \phi^4$ model. In Sec.~\ref{sec:oneloop5d} we study the one-loop order effects of \emph{one} UED on the two- and four-point vertex functions associated to the $\lambda \phi^4$ theory. In Secs.~\ref{sec:1-loopmass}  and~\ref{sec:1-looplmd} the UV divergences present in the two-point and four-point standard vertex function are respectively extracted and the counterterms to cancel them out are defined in the MS-scheme, the beta function is also presented in this subtraction scheme. In Sec.~\ref{sec:decoupling} we migrate the previous results to a mass dependent subtraction scheme and show that the decoupling theorem is satisfied, the beta function calculated in the MS-scheme is recovered when the kinematical mass scale is much larger than any mass characterising the model. In Sec.\ref{secc:conclu} the summary and final remarks are given.  By the end, the \ref{app:FR} is reserved to gather the Feynman rules of the dimensionally reduced theory. Although in the present paper we concentrate on the contribution from one UED, in a subsequent paper\citte{mar20b} we will extend our analysis to the case of the contribution from an arbitrary number $n$ of UEDs, with geometry  $\left( S^{1}/\mbb{Z}_{2}\right)^{n}$; such case requires the analysis of higher dimensional inhomogeneous Epstein zeta functions, and the Lagrangian counterterms to correctly subtract infinities will exhibit a richer structure.

\section{The dimensionally reduced scalar model from one UED}
\label{secc:5Dmodel}
In this section we describe the dimensionally reduced scalar model obtained from the self-interacting scalar field theory $\lm_{(5)}\Phi^{4}$, plus compatible operators of canonical dimension higher than five, with one UED. In the spirit of\citte{nov10, cor13, lop13, lop14} the starting point is the five dimensional field theory 
\begin{equation}\label{5Dmodel}
S=\int_{\mc{M}}\ud^{4} x \ud \bar{x}\left(\frac{1}{2}(\partial_{M}\Phi)(\partial^{M}\Phi)-\frac{1}{2}m^{2}\Phi^{2}-\frac{1}{4!}\lm_{(5)}\Phi^{4}+\sum_{r,s}\frac{\beta_{r,s}}{{\Lambda^{n+r}}}{\mathcal{O}_{s}^{(5+r)}(\Phi,\partial_{M}\Phi)}\right)\ ,\
\end{equation}
where \mbox{$\mc{M}=\mc{M}^{4}\times\left(S^{1}/\mbb{Z}_{2}\right)$}, with coordinates $(x;\bar{x})=(x^{\mu};x^{5})$, and $\Phi$ is a real valued scalar field on $\mc{M}$. The index $ M = 0,1,2,3,5 $ and Einstein sum convention for repeated indices is understood. In the scalar theory\reff{5Dmodel}, the scalar field $\Phi$ and the coupling constant $\lm_{(5)}$ have canonical dimensions of $3/2$ and $-1$, respectively; the last term contains $s$  different operators $\mathcal{O}_s^{5+r}$ of canonical dimension $5+r$, $r\geq 1$, constructed from $\Phi$ and $\partial_{\mu}\Phi$, $\Lambda$ is at an energy scale above which the new physics would begin to manifest itself, and $\beta_{r,s}$ are dimensionless parameters that depend on the details of the underlying physics. Notice that operators of higher canonical dimension are suppressed by powers of $\Lambda$. The field $ \Phi $ is assumed to fulfil the following periodicity and parity conditions
\begin{subequations}\label{PandPcond5d}
\begin{align}
\Phi(x,\bar{x}) & =\Phi(x, \bar{x}+2\pi R)\ ,  \\
\Phi(x,\bar{x}) & =\Phi(x, -\bar{x}) \ ,
\end{align}
\end{subequations}
where $R$ is the radius of the circle $S^{1}$, the latter is equivalent to imposing Neumann boundary condition at the fixed points of the orbifold\citte{che09}; relations \eqref{PandPcond5d} allow the Fourier expansion in the extra dimension of the field itself
\begin{equation}\label{5DFourier}
\Phi(x,\bar{x}) = \sqrt{\frac{1}{2\pi R}} {\phi}^{(0)}(x) + \sqrt{\frac{1}{\pi R}} \sum_{k=1}^{\infty}{\phi}^{(k)}(x)\cos\left(\frac{kx^{5}}{R}\right).
\end{equation}
As we know the coefficients in the Fourier expansion render the precise profile of the function that is being expanded, as they modulate the trigonometric functions. In this case the field modes define the profile of $\Phi$ on points in the extra dimension. It is in this sense that KK modes contain information about the behaviour of $\Phi$ on the extra dimension. Directly from the expansion~\eqref{5DFourier}, one can see that every KK mode has canonical dimension equal to 1.

Using the orthogonality of the trigonometric functions enables an immediate integration of the compact extra dimension out of the action~\eqref{5Dmodel} and defines the following dimensionally reduced theory
\begin{equation}\label{L5Dred}
\mc{L}=\mc{L}^{({0})}+\sum_{k=1}^{\infty}\mc{L}^{({0k})}+\sum_{k=1}^{\infty}\mc{L}^{(k)}+\mc{L}^{d>4}\, ;
\end{equation}
where the first term is the purely light field $  \phi^{(\unl{0})}\equiv\phi $ self-interacting $\lm\phi^{4}$ model,
\begin{subequations}\label{L5Dredterms}
\begin{equation}\label{L05Dred}
\mc{L}^{({0})}=\frac{1}{2}\left((\partial_{\mu}\phi)(\partial^{\mu}\phi)-m^{2}\phi^{2}\right)-\frac{\lm}{4!}\phi^{4}\, ,
\end{equation}
the myriad interaction terms between light and heavy fields are gathered in the second term, where
\begin{align}\label{L0k5Dred}
\mc{L}^{(0k)}=-\frac{\lm}{12}\phi\left(3\phi \phi^{(k)2}+2\phi^{(k)}\sum_{l,q=1}^{\infty}\phi^{(l)}\phi^{(q)}\Dlt_{(klq)}\right) ,
\end{align}
and the third term represents the purely KK excited mode terms, in this case,
\begin{equation}\label{Lm}
\mc{L}^{(k)}=  \frac{1}{2}\left(\partial_{\mu}\phi^{(k)}\partial^{\mu}\phi^{(k)}-m^{2}_{{(k)}}\phi^{(k)}\phi^{(k)}\right) -\frac{\lm}{4!}\phi^{(k)}\sum_{{l,q,r=1}}^{\infty}\Dlt_{({klqr})}\phi^{({l})}\phi^{({q})}\phi^{({r})}\, ,
\end{equation}
\end{subequations}
where $m^{2}_{{(k)}}=m^2+k^2/R^2$. In the set of Eqs.~\eqref{L5Dredterms}, the dimensionless multi-indexed symbols $ \Dlt_{(\cdots)} $ are defined as follows:
\begin{subequations}
\begin{align}\label{Lmdfir5d}
\Dlt_{(klq)}:=& \frac{\sqrt{2}}{\pi R}\int\ud{x^{5}}\cos\left(\frac{k{x^{5}}}{R}\right)\cos\left(\frac{l{x^{5}}}{R}\right)\cos\left(\frac{q{x^{5}}}{R}\right)\ ,\\
\Dlt_{(klqr)}:=& \frac{2}{\pi R}\int\ud{x^{5}}\cos\left(\frac{k{x^{5}}}{R}\right)\cdots\cos\left(\frac{r{x^{5}}}{R}\right)\ .
\end{align}
\end{subequations}
Directly from their definition, the multi-indexed objects $\Dlt_{(\cdots)}$ are totally symmetric. The dimensionless universal coupling constant $\lm$ is defined as $\lm_{(5)}/(2\pi R)$. The term $\mc{L}^{d>4}$ contains all the interactions of canonical dimension higher than $4$, such terms must be included in the $4$-dimensionally reduced theory because the $5$-dimensional theory is nonrenormalisable according to Dyson's criterion. It is worth mentioning that in this work we are solely interested on Green's functions at the one-loop order that only contain zero mode particles as external legs, and at most tree level effects coming from $\mc{L}_{d>4}$; remarkably, when the dimensionally reduced theory is coming from five dimensions, the counterterms needed to cancel out the UV divergences in this type of Green's functions are of canonical dimension four. 

The Feynman rules for the interactions in\reff{L5Dredterms} are gathered in the \ref{app:FR}, from them there can be classified essentially two different types of $k$-point Green's functions depending on the type of external legs, namely: the \emph{Standard Green's functions} (SGFs), whose external legs only comprise zero mode particles, and the \emph{Non-standard Green's functions} (NSGFs), whose external legs contain at least one excited KK particle. The latter can either be \emph{Hybrid Green's functions}, with some zero and some excited particles as external legs, or purely excited KK Green's functions with only KK excited particles as external legs. In particular the $k-$point SGFs, hereafter denoted by $G^{(0\ldots 0)}$ where there are $k$ slots filled with zeroes, will receive contributions from excited KK mode particles within the loops. In fact, the Lagrangian $\mc{L}^{(0)}$ together with the infinite number of interaction like the first term in\reff{L0k5Dred} provide the necessary terms to calculate the impact from the fifth dimension to $G^{(0\ldots 0)}$ at the one-loop level, in other words, to the four dimensional $\lm\phi^{4}$ theory. An interesting observation is that the superficial degree of divergence of these one-loop contributions, would indicate the possibility of removing UV divergences, that is, divergences that arise due to short distance effects in $\mc{M}^{4}$. However, as we will see later, the presence of an infinite number of interactions between the zero mode and KK modes will imply an infinite number of UV divergences at the one-loop level. This issue is characteristic of this type of theories and will be tackled on the following section.

\section{One-loop structure from one extra dimension}
\label{sec:oneloop5d}

As usual, the renormalised quantities $\lbrace \phi,\phi^{(k)},\lambda \rbrace$ and the bare quantities $\lbrace \phi_B, \phi^{(k)}_B, \lambda_B \rbrace$ are connected through the renormalisation factors as follows:
\begin{align}\label{renfactors5D}
\phi_B = \sqrt{Z}\phi, \quad \phi^{(k)}_B = \sqrt{Z_{\phi^{(k)}}}\phi^{(k)}, \quad \lambda_B = \frac{Z_{\lambda}}{Z^{2}_{\phi}}\lambda. 
\end{align}
Then, the bare Lagrangian can be written as 
\begin{equation}\label{Lbare}
\mc{L}_B = \mc{L}^{(0)}+\sum_{k=1}^{\infty}\mc{L}^{(0k)}+\sum_{k=1}^{\infty}\mc{L}^{(k)}+\mc{L}_{d>4}+\mc{L}^{(0)}_{c.t.} +\sum_{k=1}^{\infty}\mc{L}^{(k)}_{c.t.}+\mc{L}^{d>4}_{c.t.},
\end{equation}
where $\mc{L}^{(0)}$, $\mc{L}^{(0k)}$ and $\mc{L}^{(k)}$ represent the renormalised Lagrangian sectors given in the set of Eqs.\reff{L5Dredterms}, while $\mc{L}^{d>4}$ contains interactions of canonical dimension higher than four written in terms of renormalised quantities. The term $\mc{L}^{(0)}_{c.t.}$ represents the standard counterterm of the self-interacting $\lm\phi^{4}$ scalar theory, which is
\begin{align}\label{Lct05D}
\mc{L}^{(0)}_{c.t.} = \frac{1}{2}\delta_Z \left(\partial_{\mu}\phi\right)\left(\partial^{\mu}\phi\right)-\frac{1}{2}\delta{m^2}\phi^2-\frac{\delta{\lambda}}{4!}\phi^4,
\end{align}   
where 
\begin{align}
\delta_Z = Z-1, \quad  \delta{m^2} = m_B^2 Z-m^2, \quad \delta{\lambda} = \lambda_B Z^2-\lambda.
\end{align}
The contributions $\mc{L}^{(k)}_{c.t.}$ and $\mc{L}^{d>4}_{c.t.}$ in Eq.\reff{Lbare} contain interactions between the light and heavy fields, and among pure heavy fields, whose specific structure will not be needed here.

\begin{figure}
\begin{center}
\includegraphics[width=7.8cm, height=1.7cm]{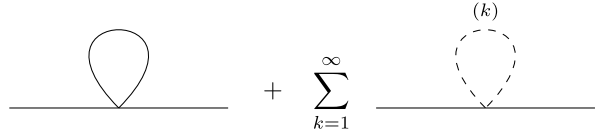}
\end{center}
\caption{{Contributions to the 2-point SVF $\Gamma_2(p)$ for one\\ extra dimension without the counterterm.}}
\label{fig:mass-self}
\end{figure}

The inverse of the 2-point connected SGF is the 2-point standard vertex function (SVF), $G_{c}^{(00)-1}=i\Gm_{2}^{(00)}$. At the one-loop order, in general, the extradimensional effects will impact the SVFs by the insertion of excited KK mode particles circulating around the loop. In the particular case of the 2-point SVF, to the 1-loop approximation, we have 
\begin{equation}
\Gamma_{2R}(p)= p^2-m^2-\left(\! M^{({0})}(p^2)+\sum_{k=1}^{\infty}M^{(k)}(p^2)+M_{c.t.}(p^2)\right),\label{2RPVF5d}
\end{equation}
where the term within parenthesis is nothing but the light scalar field self-energy which consists of three contributions: the first one, $M^{({0})}(p^2)$, comes from the well-known zero mode self-interacting term $\lm\phi^{4}$ in $ \mc{L}^{({0})} $, the second one, $\sum_{k=1}^{\infty}M^{(k)}(p^2)$, is the result of the infinite number of interactions between $\phi$ and excited KK modes present in  $ \mc{L}^{({0k})} $, henceforth we have an infinite sum of excited modes around the loop (see Fig.~\ref{fig:mass-self}), and finally the third one, $M_{c.t.}$, is the usual counterterm for the self-energy, that is,
\begin{equation}\label{SelfEnergyC}
M_{c.t.}(p^2) = \delta{m^2} - p^2\delta_Z \ ,
\end{equation}
which comes from\reff{Lct05D}. This term is sufficient to cancel out all the UV divergences present in the self-energy as it will be seen below. The first two contributions to the self-energy in Eq.\reff{2RPVF5d} are written in short as follows:
\begin{equation}\label{MmSE5d}
M^{(\boldsymbol{k})}=\frac{\lambda}{2}\int{\frac{d^4k}{(2\pi)^4}}\frac{i}{k^2-m_{{(\boldsymbol{k})}}^2}\ ,
\end{equation}

\noindent where the symbol $\boldsymbol{k}$ stands for $\lbrace 0,k \rbrace$.

\begin{figure}[h]
\begin{center}
\includegraphics[width=7.0cm, height=1.7cm]{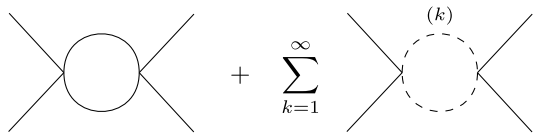}
\end{center}
\caption{Contributions to the 4-point SVF $\Gamma_4(p_i)$ without\\ the counterterm.}
\label{fig:four-vertex}
\end{figure}

If we want to analyse the corrections to the coupling constant of the $\phi^4$-theory due to the extra dimension at the one-loop, we consider the four point SVF $ \Gm_{4R}(p_{i}) $
\begin{align}\label{4RPVF5d}
\Gamma_{4R}(p_{i})=&-i\lambda+\left(\Gamma_{\trm{1-loop}}^{({0})}(p_{i})+\sum_{k=1}^{\infty}\Gamma_{\trm{1-loop}}^{({k})}(p_{i})+\Gamma_{c.t.}(p_i)\right),
\end{align}
with $\Gamma_{\trm{1-loop}}^{({0})}(p_{i})$ and $\Gamma_{\trm{1-loop}}^{({k})}(p_{i})$ the loop contributions which modify the coupling constant at low energy and $\Gamma_{c.t.}(p_i)$ the counterterm that will remove the UV infinities. The first term in the parenthesis of Eq.\reff{4RPVF5d} corresponds to the light field $\phi$ around the loop, whose presence is due to the common term $\lm\phi^{4}$ in $\mc{L}^{(0)}$, whereas the second one corresponds to an infinite sum of excited KK mode particles circulating around the loop (see Fig.~\ref{fig:four-vertex}), whose source is the infinite number of interactions between the light and heavy fields (see the first term in\reff{L0k5Dred}). Each of these terms contains a sum over the Mandelstam variables, that is,
\begin{equation}
\Gamma_{\trm{1-loop}}^{(\boldsymbol{k})}(p_{i})=\sum_{\lbrace p^2\rbrace} \Delta\Gamma^{(\boldsymbol{k})}(p)\label{Gv4pm5d}
	\end{equation}
where the symbol $\sum_{\lbrace p^{2}\rbrace}$ indicates a sum over the three Mandelstam variables, and 
	\begin{equation}
\Delta\Gamma^{(\boldsymbol{k})}(p) =-\frac{1}{2}\lambda^2\int \frac{d^4 l}{(2\pi)^4}\frac{i}{l^2-m_{{(\boldsymbol{k})}}^2}\frac{i}{(p-l)^2-m_{{(\boldsymbol{k})}}^2}.\label{DeltaGm5d}
	\end{equation}

The third term in the parenthesis in Eq.\reff{4RPVF5d} is
\begin{equation}\label{4pSVF_C}
\Gamma_{c.t.}(p_i) = -i\delta\lambda.
\end{equation}
which comes from the counterterm\reff{Lct05D}. {As we will see, this is the only counterterm needed to cancel out the divergences in\reff{4RPVF5d}. This result is remarkable, and can be rephrased by saying that the effects from one extra dimension to the self-interacting $\lm\phi^{4}$ scalar theory become finite with the aid of counterterms of canonical dimension four given by\reff{Lct05D}; however, as we will see in a subsequent paper\citte{mar20b} when more extra dimensions are involved this is no longer true, as operators of canonical dimension higher than four are needed.} 

Notice that\reff{MmSE5d} and\reff{DeltaGm5d} have identical structures to those that appear in the pure $\lm\phi^4$ theory to the one-loop, with the difference that these expressions also represent excited KK mode contributions. The above suggests that applying dimensional regularisation will reveal an infinite number of UV divergences in Eqs.\reff{2RPVF5d} and\reff{4RPVF5d} which will define the corresponding counterterms. We will show in the following subsection that this infinite number of UV divergences can be controlled by using the so-called one dimensional Epstein zeta function \citte{eps03,Ep1992, elibook12,Eli1994, eli00}.

\subsection{Regularisation}
{In order to deal with the vast number of UV divergences that at the one-loop level impact the two- and four-point SVFs, Eqs.\reff{2RPVF5d} and\reff{4RPVF5d}, respectively, we now introduce dimensional regularisation, which in turn permits to spot the emergence of the one dimensional inhomogeneous Epstein zeta function (see\citte{eps03, elibook12,Eli1994}),
\begin{equation}\label{Eps1D}
E^{c^{2}}_1(s):=\sum_{k=1}^{\infty}\frac{1}{(k^{2}+c^{2})^{s}}
\end{equation}
which becomes the Riemann zeta function when the constant $c$ vanishes.

In a scalar field theory a general vertex function contains $ N- $point scalar integrals, \ie integrals over loop momentum
\begin{equation}\label{FNmomentum}
F_{N}\propto \int\ud^{4} l\frac{1}{\prod_{j=1}^{N}\left[(l-p_{j})^{2}-m_{j}^{2}+i\veps\right]}\ ,
\end{equation}
where $m_j$ are internal masses of particles within the loop that appear in any generic vertex function to one-loop order in a given context, and $ p_{j} $ are related to the external momenta with \mbox{$ p_{1}\equiv 0 $}, \eg\reff{MmSE5d} and\reff{DeltaGm5d}. For our  divergent integrals we will apply dimensional regularisation by promoting the four dimensional spacetime to a $ D $ dimensional spacetime and define $ s:=N-D/2 $. Comparing Eq.\reff{FNmomentum} with\reff{MmSE5d} we have $N=1$ and $s\to -1$ as the four dimensional limit, whereas from\reff{FNmomentum} and\reff{DeltaGm5d} we have $N=2$ and $s\to 0$ as the four dimensional limit\footnote{At this point we decide to use the variable $s$ which is conventional in the literature of `zeta functions' (Riemann, Hurwitz, Epstein,$\ldots$), in the following section we will insert $\eps:=4-D$ which is commonly used in dimensional regularisation.}.  

{It is particularly interesting the emergence of the one dimensional inhomogeneous Epstein zeta function $E^{c^{2}}_{1}(s)$ subsequent to both dimensional regularisation and the non-cutoff of the series involving excited KK modes. Before going into that analysis, let us formally consider the contribution from all excited KK modes to the self-energy (see\reff{2RPVF5d} and\reff{MmSE5d}),
\begin{align}\label{Mkcontributionsdiscussion}
\sum_{k=1}^{\infty}M^{(k)} & =\frac{\lambda}{2}\sum_{k=1}^{\infty}\int{\frac{d^4p}{(2\pi)^4}}\frac{i}{p^2-m_{{({k})}}^2} = \frac{\lambda}{2}\sum_{k=1}^{\infty}\int{\frac{d^4p}{(2\pi)^4}}\frac{i}{\left(p^2-\dfrac{k^{2}}{R^{2}}-m^2\right)}\nonumber\\
& \equiv\frac{\lambda}{2}\sum_{k=1}^{\infty}\int{\frac{d^4p}{(2\pi)^4}}\frac{i}{\left(p_{\mu}p^{\mu}+p_{\bar{\mu}}p^{\bar{\mu}}-m^2\right)}
\end{align}
where $p^{\bar{\mu}}\equiv p^{5}= k/R$ is essentially the quantized momentum of a particle enforced by periodic conditions on the extra dimension, and $p_{\bar{\mu}}p^{\bar{\mu}}=p_{5}p^{5}=-(k/R)^{2}$ when the Lorentzian metric is used. From this viewpoint of the excited KK modes contribution, there are basically two types of potential sources of UV divergences or divergences due to small distances: one is coming from small distances in $\mc{M}^{4}$ or very large value of continuum momentum, that obviously makes the integral over $p$ diverge, and the other is coming when we allow the integer quantised momentum to increase indefinitely, in such case $p^{{5}}$ increases itself and the series over $k$ might be not well defined. Therefore one requires an analytical continuation to regularise this behaviour. In fact, when dimensional regularisation is applied such analytical continuation is actually behind the curtains justifying all integrals in $D$ dimensions as we learned from `t Hooft and Veltman\citte{thooft72}. Similar conclusion can be drawn for the excited KK modes contribution to the 4-point SVF.}

Applying dimensional regularisation to the divergent integrals in the 2-point SVF\reff{2RPVF5d}, with $ s=1-D/2 $, we obtain
\begin{equation}
\Gamma_{2R}(p^2;s) =\ p^2-m^2-\frac{\lambda}{32\pi^{2}}m^{2}\left(\frac{4\pi\mu^{2}}{m^{2}}\right)^{1+s}\left[\Gamma\left(s\right)+c^{2s}E_1^{c^2}(s)\Gamma(s)\right]-\dlt m^{2}+p^{2}\dlt_{Z},\label{2PVFReg5d}
\end{equation}
where $c^{2}:=m^{2}/R^{-2}$, $\mu$ is the common unit of mass introduced in dimensional regularisation and Eq.\reff{Eps1D} has been used. The parameter $c$ controls the scale energy of the heavy modes, $|c|\ll 1$ is the low energy limit. The first term in the squared brackets is derived from $ M^{({0})} $, this is the well-known divergent term present in the $\phi^{4}$  self-interacting scalar theory, and the second term comes from $\sum_{k=1}^{\infty}M^{({k})}$ which is the consequence of an infinite number of interactions of $ \phi $ with the excited KK modes. Since the term within the brackets in Eq.\reff{2PVFReg5d} is independent of $p^2$ we can remove $p^2\delta_Z$ by setting $\dlt_{Z}\equiv 0$; therefore the first term in the $\mc{L}_{c.t}^{(0)}$ vanishes at the one-loop order level. Needless to say that the possible UV divergences can be read as poles of either $\Gm(s)$ or the product of the one-dimensional inhomogeneous Epstein zeta function $ E^{c^{2}}_{1}(s)$ times the gamma function $\Gm(s)$, hence we must be careful in handling the limit $s\to -1$, or equivalently, $D=4$. In particular, within the product $E_1^{c^2}(s)\Gamma(s)$ one should avoid the direct application of the `limit of a product' rule. In fact, we must exhaust algebraic manipulations in this product before taking the limit. This is the main technical difference with respect to\citte{akh09}, whose analysis is performed for the case of \emph{one} spatial extra dimension $S^{1}$. If UV divergences are expected to be removed from SVFs in the low energy limit, we need to isolate divergent sources out of asymptotic formulae and judiciously define the necessary counterterms.

The same procedure of dimensional regularisation can be applied  to the 4-point SVF, leading to the expression  
\begin{align}\label{4SVFDR5d}
\Gamma_{4R}(p_{i};  s)  = & \ -i\lambda\mu^{2s}+\frac{i\lm^{2}}{32\pi^{2}}\mu^{2s}\sum_{\lbrace p^2 \rbrace} \int_{0}^{1}\ud z\left[\left(\frac{m^{2}+p^2 z(z-1)}{4\pi\mu^{2}}\right)^{-s}\Gm(s)\right.\notag\\
& \left.\ +\left(\frac{4\pi \mu^{2}}{R^{-2}}\right)^{s}E^{c(p,z)^{2}}_{1}(s)\Gm(s)\right]-i\delta\lambda,
\end{align}
being $c(p,z)^{2}:=(m^{2}+p^2 z(z-1))/R^{-2}$. The first term in the squared brackets represents the correction to the coupling constant coming from the light field interacting term $\lm\phi^4$, this is, $\Gamma_{\trm{1-loop}}^{({0})}(p_{i})$; while the second term comes from the extra-dimensional contribution $\sum_{k=1}^{\infty}\Gamma_{\trm{1-loop}}^{({k})}(p_{i})$ and $\delta\lambda$, as previously stated, defines the corresponding counterterm. 
    
\section{One-loop mass corrections in the presence of one extra dimension}
\label{sec:1-loopmass}
The counterterms defined in these subsections in order to deal with the one-loop effects from the fifth dimension at the level of the mass and coupling constant of the self-interacting scalar field theory $\lm\phi^{4}$ will be inspired by the MS-scheme\citte{colbook, tho78}, \ie a mass independent scheme at which the UV divergences of Feynman graphs will be cancelled out by terms defined as poles at $D=4$. Interestingly, as we have mentioned such terms define the countertemrs of canonical dimension four, see\reff{Lct05D}.

In the presence of a single extra dimension, we analyse the UV divergences in $\Gm_{2R}$ and $\Gm_{4R}$. The non-trivial source of these poles is found in $E^{c^{2}}_{1}\Gm $ in the corresponding four-dimensional limit (see Eqs.\reff{2PVFReg5d} and\reff{4SVFDR5d}). Our problem then comes down to investigate the structure of the poles of $ E^{c^{2}}_{1}\Gm$ and this will be done in the limit $\vert c \vert \ll 1$, where the formula for this product is known, see\citte{mar20b} or\citte{Ep1992}\footnote{We must say that in the sense of\citte{Ep1992} this expression is nothing but a \emph{non}-regularised version of the product $E_{1}^{c^{2}}(s)\Gm(s)$.}, namely
\begin{equation}
E_1^{c^2}(s)\Gamma(s)=\sum_{k=0}^{\infty}\frac{(-1)^k}{k!}\Gamma(k+s)\zeta(2k+2s)c^{2k}\ .\label{AEa1}
\end{equation}

As it can be seen from\reff{AEa1}, the 2-point SVF can be written down as follows in the low energy limit, 
\begin{align}\label{AsymGen5d}
\Gm_{2R} (p^{2},\eps) & =  p^{2}-m^{2}-\frac{\lm m^{2}}{32\pi^{2}}\left(\frac{4\pi\mu^{2}}{m^{2}}\right)^{\eps/2}\bigg[\Gm\left(\frac{\eps}{2}-1\right)\nonumber\\
&\ \ +\sum_{k=0}^{\infty}\frac{(-1)^{k}}{k!}{\Gm\left(\frac{\eps}{2}-1+k\right)} \zeta\left(2k+\eps-2\right)c^{2k+\eps-2}\bigg]-\delta m^2\, ,
\end{align}
where the four dimensional limit is obtained when $\epsilon\rightarrow 0$, with $\epsilon/2 \equiv s+1= 2-D/2$, that is, $s\rightarrow -1$.   
The first term in the squared brackets contributes with the well-known UV divergence, namely the one resulting from the following asymptotic expansion,\footnote{We use the twiddle sign $\sim$ to mean asymptotic equalities.}
\begin{align}\label{Gm-asymp5D}
\left(\frac{4\pi\mu^{2}}{m^{2}}\right)^{\eps/2}\Gm\left(\frac{\eps}{2}-1\right)\sim -\frac{2}{\epsilon}+\text{(finite)}+\mc{O}(\eps) ,
\end{align}
here, the `(finite)' part contains typical constants like the Euler-Mascheroni $\gamma_E$ and a logarithm that includes in its argument the ratio $4\pi\mu^2/m^2$. The analysis on the second term in\reff{AsymGen5d} is subtler, the possible sources of divergences may come either from the arguments of the gamma or the Riemann zeta function $\zeta$. Therefore, the poles will be found when the argument of the gamma function is either a negative integer or zero, and/or when the argument of the Riemann zeta function becomes one. These conditions can be expressed as follows:
\begin{subequations}
\begin{align}
-2k+2& =2j\ , \quad j\in\mbb{N}_{0}\label{rgama5d}\\
-2k+2 & =-1\ .\label{rzeta5d}
\end{align}
\end{subequations}
The Eq.\reff{rgama5d} is only satisfied for $k=0$ and $k=1$, and there is no positive integer $k$ that fullfils the identity\reff{rzeta5d}. Therefore the only source of UV divergences within $\Gm_{{2R}}$ for one extra dimension is encoded in the gamma function, this will not be the case for greater number of extra dimensions\citte{mar20b}. At $k=0$ the relevant asymptotic expression as $\epsilon\rightarrow 0$ is the following
\begin{align}\label{Asympk05d}
\left(\frac{4\pi \mu^2 c^2}{m^2}\right)^{\epsilon/2}\Gamma(\epsilon/2-1)\zeta(\epsilon-2)\sim& 
\frac{\zeta(3)}{2\pi^2} +\text{(finite)}+\mathcal{O}(\epsilon).
\end{align} 
At $k=1$ the relevant asymptotic expression for small $\epsilon$ is
\begin{align}\label{Asympk15d}
\left(\frac{4\pi \mu^2 c^2}{m^2}\right)^{\epsilon/2}\Gamma(\frac{\epsilon}{2})\zeta(\epsilon)\sim &
\frac{2\zeta(0)}{\epsilon}+\text{(finite)}+\mathcal{O}(\epsilon).
\end{align}
From the relation\reff{Asympk05d} notice that although the gamma function diverges as $\epsilon$ goes to zero, the product $\Gamma(\epsilon/2-1)\zeta(\epsilon-2)$ is finite. In other words, the term corresponding to $k=0$ does not contribute to the pole at all. This is not the case for the term $k=1$, as we can see from the asymptotic equality\reff{Asympk15d} which shows the typical ultraviolet divergence $1/\epsilon$ as $\epsilon\rightarrow 0$. 

{Introducing the asymptotic relations\reff{Gm-asymp5D},\reff{Asympk05d} and\reff{Asympk15d} into the 2-point SVF\reff{AsymGen5d} we have for small $\eps$
\begin{align}\label{Gm2expanded5D}
\Gm_{2R}  (p^{2} & ,\eps) =   p^{2}-m^{2}-\frac{\lm m^{2}}{32\pi^{2}}\sum_{k=2}^{\infty}\frac{(-1)^{k}}{k!}{\Gm\left(k-1\right)} \zeta\left(2k-2\right)c^{2k-2}\nonumber\\
& +\left[\frac{\lambda m^2}{16\pi^2}\left(\frac{1}{\epsilon}\right)+\frac{\lambda m^2\zeta(0)}{16\pi^2}\left(\frac{1}{\epsilon}\right)\right]+\frac{\lm m^{2}\zeta(3)}{2\pi^{2}c^{2}}+(\!\!\trm{finite}) +\mc{O}(\eps)-\delta m^2\, .
\end{align}
Since the series left after isolating the divergent terms (at $k=0$ and $k=1$) converges in the limit $\vert c \vert\ll 1$, we conclude that the 2-point SVF\reff{2PVFReg5d} will become UV finite at the one-loop level if we define the proper counterterm to avoid the ultraviolet divergence that we just found. In the spirit of the MS-scheme prescription \cite{colbook,tho78}, where the UV divergences of a Feynman graph must be cancelled out by counterterms that can be read from the poles at $D=4$, we define
\begin{equation}\label{counterterm5dM}
\delta m^2:=\frac{\lambda m^2}{16\pi^2}\left(\frac{1}{\epsilon}\right)+\frac{\lambda m^2\zeta(0)}{16\pi^2}\left(\frac{1}{\epsilon}\right).
\end{equation}
and hence the second term in $\mc{L}_{c.t.}^{(0)}$\reff{Lct05D} becomes totally defined. On the right hand side of Eq.\reff{counterterm5dM} the first term is the typical pole in the self-energy coming from $\mathcal{L}^{(0)}$, and the second term is the extra dimensional or KK contribution. The presence of the factor $\zeta(0)=-1/2$ is the regularised version of the foreseen UV divergence associated to small distances in the extra dimension\footnote{Remember that the Euler-zeta function $\zeta_{E}(s)=\sum_{k=1}^{\infty}\frac{1}{n^{s}}$ diverges in the Cauchy sense at $s=0$, however once this function has been analytically continued into the Riemann zeta function $\zeta(s)$ it has well defined value.} 

When the heavy fields become infinitely massive, that is, in the limit $c\rightarrow 0$ or equivalently $R^{-1}\rightarrow \infty$, we expect in Eq.\reff{Gm2expanded5D} the decoupling of the KK contribution, however this does not happen, notice the term proportional to $1/c^{2}$ which diverges in the limit $c\rightarrow 0$; even more, the `(finite)' part in this expression contains a logarithmic divergent term proportional to $\log (2\pi c)$. Therefore, although the MS-scheme removes the UV divergences, it is not concerned about the decoupling theorem; nevertheless a mass dependent scheme to cancel out divergences ensures the decoupling as we will show in Sec. \ref{secc:Decoupling5D}. This is also true in the dimensionally reduced version of QED as it was shown in\citte{QED5d}.

\section{One-loop corrections to the coupling constant in the presence of one extra dimension}
\label{sec:1-looplmd}

We now apply a similar treatment to $\Gm_{4R}(p;s)$ given in Eq.\reff{4SVFDR5d}. This expression diverges as $s\rightarrow 0$, or equivalently as $\epsilon\rightarrow 0$ with $\epsilon/2\equiv s $. Since our interest is to isolate UV divergences, the idea is then to examine the pole structure of the $\Gamma_{4R}(p_i;s)$ and for that we will take advantage of the expression in the low energy limit, $|c|\ll 1$, Eq.\reff{AEa1}. Then the 4-point SVF can be expressed as  
\begin{align}\label{4SVFc5d}
 \Gamma_{4R}(p_{i} & ;\eps)= -i\lambda\mu^{\eps}+\frac{i\lm^{2}\mu^{\eps}}{32\pi^{2}}\sum_{\lbrace p^{2} \rbrace} \Bigg[\Gm\left(\frac{\eps}{2}\right) \int_{0}^{1}\ud z\left(\frac{m^{2}+p^{2}z(z-1)}{4\pi\mu^{2}}\right)^{-\eps/2}\! \notag\\
& +\sum_{k=0}^{\infty}\frac{(-)^{k}}{k!}\left(\frac{4\pi \mu^{2}}{m^{2}}\right)^{{\eps/2}}\!\!\Gm\left(\frac{\eps}{2}+k\right) \zeta(\eps+2k)F_{k}(p^{2})c^{\eps+2k}\Bigg]-i\delta\lambda 
\end{align}
where $F_{k}(p^{2})$ is defined as 
\begin{align}
F_{k}(p^{2}) :=& \left( \frac{4\pi\mu^{2}}{m^{2}}\right)^{k}\int_{0}^{1}\ud z\left(\frac{m^{2}+p^{2}z(z-1)}{4\pi\mu^{2}}\right)^{k}\! \notag\\
=& \sum_{l=0}^{k}\binom{k}{l}\frac{(-1)^{l}(l!)^{2}}{(2l+1)!}\left(\frac{p^{2}}{m^{2}}\right)^{l} .
\end{align}

The contribution to $\Gamma_{4R}(p_i;s)$ from the light particle around the loop is encoded in the first term in the squared brackets of \reff{4SVFc5d}. This is the well-known contribution from $\mathcal{L}^{(0)}$, the corresponding input to the pole can be calculated using
\begin{align}\label{4SVF-0cont}
\Gm\left(\frac{\epsilon}{2}\right)\left(\frac{m^{2}+p^2 z(z-1)}{4\pi\mu^{2}}\right)^{-\epsilon/2}
\thicksim \frac{2}{\epsilon}+\text{(finite)}+\mathcal{O}(\epsilon),
\end{align}
where the `(finite)' part in\reff{4SVF-0cont} contains the Euler-Mascheroni constant and a logarithm which depends on momentum as $(m^{2}+p^2 z(z-1))/4\pi \mu^{2}$ is integrated in the $z$ variable.
Regarding the second term in the squared brackets of\reff{4SVFc5d},
the divergent terms can be known from the analysis of the arguments of the gamma and Riemann zeta functions, in fact, these functions diverge whenever
\begin{subequations}
\begin{align}
-2k& =2j, \quad j\in\mbb{N}_{0}\label{rgama25d}\\
-2k & =-1,\label{rzeta25d}
\end{align}
\end{subequations}
respectively. In this case, there is only one solution for\reff{rgama25d} that is when $k=0$, and there is no solution for\reff{rzeta25d} since there is no positive integer $k$ that fullfils it. Hence, the only source of UV divergence at $\Gm_{4R}$ comes also from the gamma function only. When $k=0$ the asymptotic of such term can be obtained from\reff{Asympk15d}. {Introducing the asymptotic behaviour relations\reff{4SVF-0cont} and\reff{Asympk15d} into\reff{AsymGen5d}, we get for small~$\eps$ 
\begin{align}\label{4SVFc5dsmalleps}
\Gamma_{4R}(p_{i};\eps)=& \ -i\lambda\mu^{\eps}+\frac{i\lm^{2}\mu^{\eps}}{32\pi^{2}}\sum_{\lbrace p^{2} \rbrace}\sum_{k=1}^{\infty}\frac{(-)^{k}}{k!}\Gm\left(k\right) \zeta(2k)F_{k}(p^{2})c^{2k}\notag\\
&+\Bigg[\frac{3i\lm^{2}}{16\pi^{2}}\left(\frac{\mu^{\eps}}{\eps}\right)+ \frac{3i\lm^{2}\zeta(0)}{16\pi^{2}}\left(\frac{\mu^{\eps}}{\eps}\right)\Bigg]+(\!\!\trm{finite})+\mc{O}(\eps)-i\delta\lambda 
\end{align}
from which the counterterm $\dlt\lm$ can be established, so that the finite version of the 4-point SVF, at the one-loop level, is obtained by defining
\begin{align}\label{CounterLambda5d}
i\delta \lambda := \frac{3i\lambda^2 }{16\pi^2}\left(\frac{1}{\epsilon}\right)+\frac{3i\lambda^2 \zeta(0)}{16\pi^2}\left(\frac{1}{\epsilon}\right).
\end{align}
With this particular value of $\dlt\lm$ the last counterterm in $\mc{L}_{c.t.}$ becomes completely defined. On the right hand side of\reff{CounterLambda5d}, the first term is the typical UV divergence in the $\phi^4$ theory $\mathcal{L}^{(0)}$, and the second term is the UV divergence coming from the extra dimension at the one-loop level in the low energy limit, which is again characterized by the regularised value $\zeta(0)$; the factor $3$ comes from the sum $\sum_{\lbrace p^2 \rbrace}$. } In connection with the extension of this result to the presence of higher number of UEDs, it is essential to stress that\reff{CounterLambda5d} is momentum independent, the reason being that the UV divergence comes from the term $k=0$ in\reff{4SVFc5d} which contains the factor $F_{0}(p^{2})=1$; when more extra dimensions are compactified, the resulting $\Gm_{4R}$ of theory contains UV divergences within terms that involve genuine functions of momentum $F_{k}(p^{2})$, this fact implies a richer structure in the related counterterms\citte{mar20b}.

Notice that the needed counterterms to cancel out UV divergences present in the SVFs $\Gm_{2R}$ and $\Gm_{4R}$ are of canonical dimension four, that is, we do not appeal to any interaction included in $\mc{L}_{c.t.}^{d>4}$; however, this is a very special case since in the presence of more extra dimensions there will be needed counterterms of canonical dimension higher than four and interactions in $\mc{L}_{c.t.}^{d>4}$ will become relevant\citte{mar20b}. 

Directly from the 4-point SVF, the beta function $\beta(\lambda)$ can be calculated at the one-loop order, in this case,
\begin{align}\label{beta5d}
\beta(\lambda)& =\lim_{\epsilon\rightarrow 0}\mu\frac{\partial \lambda}{\partial\mu}=\lim_{\epsilon\rightarrow 0}\mu\frac{\partial}{\partial\mu}\left(\lambda\mu^{\epsilon}+ \frac{3i\lambda^2 }{16\pi^2}\left(\frac{\mu^{\epsilon}}{\epsilon}\right)+\frac{3i\lambda^2 \zeta(0)}{16\pi^2}\left(\frac{\mu^{\epsilon}}{\epsilon}\right)\right)\notag\\
& =\frac{3\lambda^2}{16\pi^2}+\frac{3i\lambda^2 \zeta(0)}{16\pi^2}\\
& =\frac{3\lambda^2}{32\pi^2}.
\end{align}
The beta function measures the strength of the coupling constant with energy. On the right hand side of Eq.\reff{beta5d} the first term is the well-known value of the beta function for the self-interacting $\phi^4$ theory in the MS-scheme, whereas the second term is the contribution due to the fifth dimension. The value of $\beta(\lambda)$ is positive, which means that the strength of the coupling increases as the energy increases. Notice that the extra dimensional contribution reduces by a factor $1/2$ the beta function in the absence of the fifth dimension. 

It is important to emphasize the fact that when $R^{-1}\rightarrow \infty$ ($c\rightarrow 0$) the decoupling theorem is not fulfilled either for the $\Gm_{4R}$ or the beta function. Indeed, regarding $\Gm_{4R}$, see Eq.\reff{4SVFc5dsmalleps}, one must say that inside the `(finite)' part there is a term proportional to $\log (2\pi c)$ which makes the decoupling theorem not evident. In addition, the beta function is simply an $R$-independent constant, therefore the decoupling theorem is not manifest at all. This is not a surprise since as it is well known: the major difference between dimensional regularisation together with the MS-scheme and other schemes is that the $\bt$ function is independent of the scale\citte{QED5d, Petrov}. In order to restore scale dependence we should migrate to a mass dependent subtraction scheme, this will be the main idea in the following section.

\section{Decoupling of the KK contribution in the presence of one extra dimension}\label{secc:Decoupling5D}
\label{sec:decoupling}

In this section we will migrate to a mass dependent subtraction scheme by the choice $p^2=-M^2$, with $M$ the kinematical or subtraction scale. We begin by analysing the one-loop contribution of the self-energy given by (see\reff{2PVFReg5d} with $s+1=\epsilon/2$)
\begin{equation}
M_{5D}(p^2):=  \frac{\lambda}{32\pi^{2}}m^{2}\left(\frac{4\pi\mu^{2}}{m^{2}}\right)^{\epsilon}\left[\Gamma\left(\frac{\epsilon}{2}-1\right) +\, E_1^{c^2}(\frac{\epsilon}{2}-1)\Gamma(\frac{\epsilon}{2}-1)c^{\eps-2}\right]-\dlt m^{2}+p^{2}\dlt_{Z},
\end{equation}    
where none of the terms in the squared brackets depend on $p^{2}$, hence $\dlt_{Z}=0$. We determine the counterterm $\delta m^2$ using the typical kinematical condition
\begin{equation}\label{rencond5D}
M_{5D}(p^2)_{\big{|}_{p^2 = -M^2}}=0\, ,
\end{equation} 
therefore
\begin{align}
\delta m^2 = & -\frac{\lambda m^2}{32\pi^2}\left(\frac{4\pi \mu^2}{m^2}\right)^{\epsilon/2}\Gamma\left(\frac{\epsilon}{2}-1\right)\left[1+E_1^{c^2}\left(\frac{\epsilon}{2}-1\right)c^{\epsilon-2}\right].
\end{align}
In this scheme the self-energy $M_{5D}(p^2)=0$ for all $p^2$. At the one-loop level, the renormalised mass is not impacted by extra dimensions once the UV divergences are cancelled out, this is because $M_{5D}(p^2)$ does not depend on the external momentum $p^2$ unlike in the dimensionally reduced QED\citte{QED5d}. At this level, not much can be said about the decoupling, and we might have to extend our calculations to two loops if we want to explicitly see the decoupling of the KK contribution. Still, the  4-point SVF is more enlightening in this aspect.

For the 4-point SVF we have an explicit dependence on the external momenta at the one-loop order. The 4-point SVF\reff{4SVFDR5d} when $\eps/2=s$ is introduced becomes,
\begin{align}\label{Gm4R5Dmassdependent}
\Gamma_{4R\,{5D}}(p_i,\epsilon) =& -i\lambda \mu^{\epsilon} + \frac{i\lambda^2}{32\pi^2}\mu^{\epsilon}\sum_{\lbrace p^2 \rbrace}\int^1_0 dz \left[\left(\frac{m^2+p^2z(z-1)}{4\pi \mu^2}\right)^{-\epsilon/2}\Gamma(\epsilon/2)  \right.\nonumber\\
&\left. \, +\, \sum_{k=1}^{\infty}\left(\frac{m^2_{(k)}+p^2z(z-1)}{4\pi \mu^2}\right)^{-\epsilon/2} \Gamma(\epsilon/2)\right]-i\delta{\lambda}.
\end{align}
In order to determine $\delta{\lambda}$, we appeal to the following condition:
\begin{align}\label{4svfrencon5D}
\Gamma_{4R\,{5D}}(s,t,u = -M^2,\epsilon) = -i\lambda\mu^{\epsilon}.
\end{align}
then,
\begin{align}\label{DLC5}
i\delta{\lambda} & =\frac{3i\lambda^2}{32\pi^2}\mu^{\epsilon}\int^1_0 dz \Gamma(\epsilon/2)\left[\left(\frac{m^2-M^2z(z-1)}{4\pi \mu^2}\right)^{-\epsilon/2}+ \sum_{k=1}^{\infty}\left(\frac{m^2_{(k)}-M^2z(z-1)}{4\pi \mu^2}\right)^{-\epsilon/2}\right], 
\end{align}
using the asymptotic expression\reff{4SVF-0cont}, we have for small $\eps$
\begin{align}\label{lm1massdependent}
i\delta{\lambda} & =  \frac{3i\lambda^2}{32\pi^2}\int_0^1 dz \Bigg{\lbrace}\frac{2}{\epsilon}-\gamma_E + \log(4\pi) - \log\left(\frac{m^2-z(z-1)M^2}{\mu^2}\right)\nonumber \\
&+\sum_{k=1}^{\infty}\left[\frac{2}{\epsilon}-\gamma_E + \log(4\pi)- \log\left(\frac{m_{(k)}^2-z(z-1)M^2}{\mu^2}\right) \right]\Bigg{\rbrace}+\mc{O}(\eps),
\end{align} 
where we have explicitly written all the `(finite)' terms.
Then the UV-divergent free 4-point SVF is obtained by the substitution of\reff{lm1massdependent} into\reff{Gm4R5Dmassdependent} giving
\begin{align}\label{MOSG4}
\Gamma_{4_{5D}}(p_i) =& -i\lambda-\frac{i\lambda^2}{32\pi^2}\int^1_0 dz \sum_{\lbrace p^2 \rbrace}\left[\log\left(\frac{m^2+z(z-1)p^2}{m^2-z(z-1)M^2}\right)\right.\nonumber\\
&\left.+\sum_{k=1}^{\infty}\log\left(\frac{m_{(k)}^2+z(z-1)p^2}{m_{(k)}^2-z(z-1)M^2}\right)\right].
\end{align}
The decoupling theorem is now manifest in Eq.\reff{MOSG4}, when the mass of the heavy modes is much larger than our kinematical mass, $m_{(k)}\gg M$, we have
\begin{align}
\Gamma_{4R\,{5D}}(p_i) =  & -i\lambda-\frac{i\lambda^2}{32\pi^2}\int^1_0\!\! dz \sum_{\lbrace p^2 \rbrace}
\log\left(\frac{m^2+z(z-1)p^2}{m^2-z(z-1)M^2}\right)\, ,
\end{align}
then only the contribution of the light field is present. Even more, the $\beta$ function can be calculated in this mass dependent scheme using\citte{Petrov}
\begin{align}
\beta(\lambda) = M\frac{\partial \delta\lambda_1}{\partial M}=\frac{3\lambda^2}{16\pi^2}\int_0^1 dz \left[\frac{z(1-z)M^2}{m^2+z(1-z)M^2}+\sum_{k=1}^{\infty}\frac{z(1-z)M^2}{m_{(k)}^2+z(1-z)M^2}\right],
\end{align}\label{betaMOS}
in particular, when $m\ll M$, the $\beta$ function approaches to 
\begin{equation}
\beta(\lambda) = \frac{3\lambda^2}{16\pi^2}\left[1+\int_{0}^{1}\ud z\sum_{k=1}^{\infty}\frac{z(1-z)M^2}{m_{(k)}^2+z(1-z)M^2}\right]\ .
\end{equation}
In addition, in the limit $ M \ll m_{(k)}$, we recover the well-known value for the self-interacting scalar theory, since all KK contributions decouple, leading to the value
\begin{align}
\beta(\lambda) = \frac{3\lambda^2}{16\pi^2}.
\end{align}
If on the other hand, we permit $M$ to be extremely large, \ie $M\gg m_{(k)}$ and $M\gg m$, simultaneously, the value for the $\beta$ function formally becomes the one obtained in the MS-scheme\reff{beta5d}, that is,
\begin{align}
\beta_{5D}(\lambda) =& \frac{3\lambda^2}{16\pi^2}\left[1+\int_{0}^{1}\ud z\sum_{k=1}^{\infty}\frac{z(1-z)M^2}{m_{(k)}^2+z(1-z)M^2}\right]=  \frac{3\lambda^2}{16\pi^2}\left[1+\int_{0}^{1}\ud z\, \zeta(0)\right]\nonumber\\
 =&\frac{3\lambda^2}{16\pi^2}+\frac{3\lambda^2\zeta(0)}{16\pi^2}.
\end{align} 

\section{Summary and final remarks}
\label{secc:conclu}

In this paper we have presented the one-loop order contributions from one extra dimension to the self-energy and 4-point vertex functions associated to the $\lm\phi^4$ self-interacting scalar theory in the context of a UED. We explore such contributions employing the dimensionally reduced Lagrangian obtained from a  field theory defined on the five dimensional spacetime  $\mc{M} ^{4}\times S^{1}/Z_{2}$; this theory corresponds to the effective Lagrangian formed by the self-interacting $\lm_{(5)}\Phi^{4}$ model plus all of the Lorentz invariant operators of canonical dimension higher that five Eq.\reff{5Dmodel}. The structure of the dimensionally reduced theory, defined on the Minkowski spacetime $\mc{M}^{4}$, is given by the following sectors: $(i)$ the pure $\lm\phi^{4}$ theory, with $\phi$ the lightest field in the Kaluza-Klein tower that comes naturally after compactification, $(ii)$ a sector that has an infinite number of interactions between $\phi$ and the excited KK fields, $(iii)$ a sector with a kinematical term and interactions of purely KK excited fields $\phi^{(k)}$, and $(iv)$  a sector that comes from the dimensional reduction of the Lorentz invariant operators of canonical dimension higher than five. We suggested that dealing with sectors $(i)$ and $(ii)$, together with the necessary counterterms, allows to investigate on the impact of the fifth dimension to the self-energy and 4-point vertex functions of the $\lm\phi^{4}$ model. 

Regarding the self-energy, we show  that at the one-loop order, besides the usual scalar particle around the loop, there appears a contribution given by an infinite superposition of KK excited modes. This contribution, which before regularisation seems to contain an infinite number of UV divergences, can  analytically be recognised with the one dimensional inhomogeneous Epstein zeta function times the gamma function once dimensional regularisation is performed. In fact, this product can also be written as a series of products between the gamma and the Riemann zeta functions from which one is able to extract the term responsible for the UV divergence coming from the extra-dimensional contribution. The emergent UV divergence is proportional to $\zeta(0)/\eps$, where $\eps\to 0$  when we approach to four dimensions. This term together with the typical divergence due to the light particle around the loop were cancel out by a Lagrangian counterterm of canonical dimension four of the usual structure $-\dfrac{1}{2}\dlt m^{2}\phi^{2}$. In other words, there was no need to rely on operators of canonical dimension higher that four to cancel out all the UV divergences present at $\Gm_{2R}$ to the one-loop order. The UV divergences were removed using two different subtraction schemes: the MS-scheme and a mass dependent subtraction scheme. In none of them the decoupling theorem is manifest, although the latter scheme may show this property if we go to a two-loop order in the expansion.

We also investigate the impact of the fifth dimension to the 4-point vertex function of the $\lm\phi^{4}$ model. Again, it was proved that at the one-loop order, besides the usual light particle around the loop there exists a contribution formed by an infinite number of  excited KK particles. Therefore one has to deal with a seemingly infinite number of UV divergences at a finite order in the perturbative expansion. Using dimensional regularisation we show that the infinite number of these divergences can be encapsulated, again, into the product of the one dimensional inhomogeneous Epstein zeta function times the gamma function. As we previously mention, this product finds a more tractable form in the low energy limit, where it can be expressed as a series of products between the gamma and the Riemann zeta functions, so that extracting the UV divergences comes down to localise poles of these products. As in the case of the self-energy, it was proved that besides the UV divergence coming from the light particle around the loop, the extra-dimensional contribution contains one UV divergence which behaves as $\zeta(0)/\eps$. Hence in order to cancel them out only Lagrangian counterterms of canonical dimension four are needed, in fact they can be gathered into $-\dfrac{\dlt\lm}{4!}\phi^{4}$. Again we use the MS-scheme and a mass dependent subtraction scheme with a kinematical mass denoted by $M$ to cancel out divergences. Although the former is not concerned about the decoupling theorem, the latter explicitly shows that the theorem is manifestly satisfied at the level of the $\Gm_{4R}$ and, more importantly, at the level of the beta function. In this regard, the beta function associated to the $\lm\phi^{4}$ pure model is recovered in the limit of excited KK modes much heavier than $M$. Moreover, the beta function calculated in the MS-scheme is recovered in the case where the kinematical mass is much heavier than the heavy and light fields. All the aforementioned results will be extended in a forthcoming paper\citte{mar20b} to the case where more than one extra dimension is present. 

It becomes stimulating to realize that the whole dimensionally reduced model still contains ingredients which could be the source of some theoretically interesting investigations. Besides the problem to systematize the study of the complete zoo of NSGFs at the one-loop level, the term proportional to $-\lm\phi\sum_{{kln}}\phi^{({k})}\phi^{({l})}\phi^{({n})}\Dlt_{({kln})}$ in the sector\reff{L0k5Dred} contains, once the multi-indexed $\Dlt_{({kln})}$ is explicitly written, the term proportional to $-\lm\phi\sum_{k}\phi^{(2{k})}\phi^{({k})}\phi^{({k})}$. Such term would give rise to the presence of an infinite number of $\phi -\phi^{(2{k})}$ transitions at the one-loop level with a $\phi^{({k})}$ particle within the loop (see the Feynman rule\reff{fig:Feyn3}), in addition there will also be an infinite number of mixings among excited KK modes; these effects resemble the interesting $\gm$-$Z$ mixing in the Weinberg-Salam model which has been studied in linear and non-linear $R_{\xi}$ gauges\citte{bau82, rom87} and whose resolution relies on a particular gauge. 

\section*{Acknowledgements}
We acknowledge financial support from CONACyT (Mexico). M.A.L.-O, E.M.-P. and J.J.T. acknowledge SNI (Mexico). 

\appendix
\section{Feynman rules for the dimensionally reduced theory}
\label{app:FR}
In this section we gathered the Feynman rules for the dimensionally reduced theory coming from five dimensions. From the zero mode sector $\mc{L}^{(0)}$ we have the usual Feynman rules of the self-interacting $\phi^{4}$ theory, in particular the interaction term gives
\begin{equation}\label{fig:Feyn1}
\begin{matrix}
\includegraphics[width=130.0 pt]{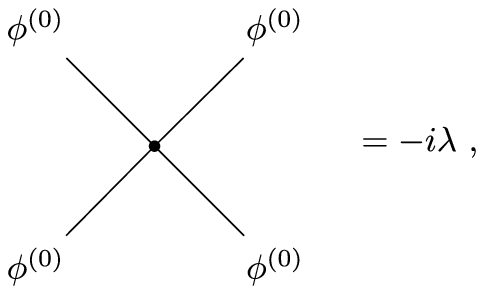} 
\end{matrix}
\end{equation}
from the interaction terms contained in $\mc{L}^{(0k)}$ we have the following:
\begin{equation}\label{fig:Feyn2}
\begin{matrix}
\includegraphics[width=130.0 pt]{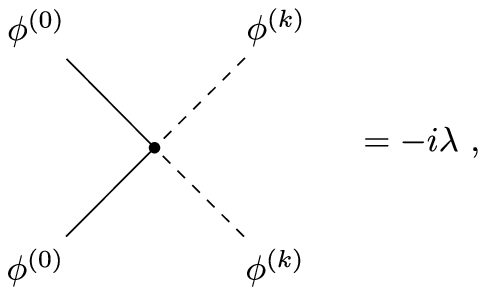} 
\end{matrix}
\end{equation}

\begin{equation}\label{fig:Feyn3}
\begin{matrix}
\includegraphics[width=150.0 pt]{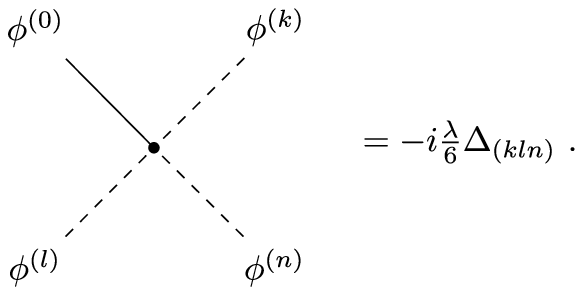} 
\end{matrix}
\end{equation}
Finally, from the purely excited KK sector $\mc{L}^{(k)}$  we read the following interaction rule:
\begin{equation}\label{fig:Feyn4}
\begin{matrix}
\includegraphics[width=150.0 pt]{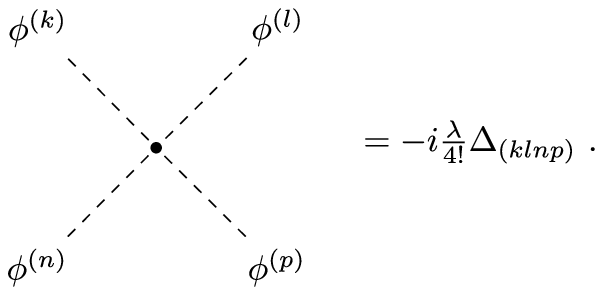} 
\end{matrix}
\end{equation}

Notice that in fact, from the rule\reff{fig:Feyn3} we have a mixing phenomenon between $\phi$ and $\phi^{(2k)}$ that mimics the well-known $\gm$-$Z$ mixing present in the Weinberg-Salam model\citte{bau82, rom87}.

\end{document}